\documentstyle[prb,preprint,aps,eqsecnum]{revtex}
\tighten
\begin{document}
\draft
\title{RELATIVISTIC PARTICLE IN THE LIOUVILLE FIELD}
\author{George Jorjadze\\}
\address{Department of Theoretical Physics, Razmadze Mathematical
Institute, Tbilisi, Georgia.\\E-mail: jorj@imath.acnet.ge\\}
\author{W{\l}odzimierz Piechocki\\}
\address{Field Theory Group, So{\l}tan Institute for Nuclear 
Studies, Warsaw, Poland.\\E-mail: piech@fuw.edu.pl \\}
\date{\today}
\maketitle
\begin{abstract}
A model of a relativistic particle moving in the Liouville field is 
investigated. Symmetry group of the system is $SL(2,R)/Z_2$. 
 The corresponding dynamical integrals describe full set of 
classical trajectories. Dynamical integrals are used for the 
gauge-invariant
 Hamiltonian reduction. The new scheme is proposed 
for quantization of the reduced system. Obtained quantum system 
reproduces classical symmetry. Physical aspects of the model are 
discussed.
\end{abstract}

\vspace{10mm}


\section{Introduction}
In 2-dimensions the gravitational field $g_{\mu \nu}(X)$
($X:=(x^0,x^1)$; 
$(\mu ,\nu = 0,1))$
can be characterized only by a single field $\varphi (X)$, since
one can always choose coordinates in such a way that $g_{\mu \nu}$
takes the form (at least locally) \cite{1}
\begin{equation}
 g_{\mu \nu}(X) = \exp{\varphi (X)}\;\;\left( \begin{array}{cr}
 1 & 0 \\ 0 & -1
 \end{array} \right).
 \end{equation}
The scalar curvature $R(X)$ calculated for the metric tensor (1.1) 
is given by
\begin{equation}
R(X)=e^{- \varphi (X)}(\partial ^2_1- \partial ^2_0)\varphi (X).
\end{equation}
Therefore, the Einstein-Hilbert Lagrangian 
${\cal {L}}=-\sqrt{\mid g(X)\mid}R(X)$  
degenerates in 2-dimensions 
(${\cal {L}}=(\partial ^2_0- \partial ^2_1)\varphi (X)$)
and it does not lead to dynamical equation for the field $\varphi (X)$.

One can specify the 2d general relativity model by requiring that the 
field $\varphi$ satisfies a dynamical equation.
The Liouville field equation
\begin{equation}
(\partial ^2_0 - \partial ^2_1)\;\varphi (x^0,x^1) + R_0 \exp{\varphi 
(x^0,x^1)} = 0
\end{equation}
(where $R_0$ is a non-zero constant) is usually considered
as a model of such a theory\cite{2,3}. The field $\varphi (x^0,x^1)$
satisfying (1.3) leads to the manifold with a constant curvature $R(X)
=R_0\;\;$ (see (1.2)). Eq.(1.3) 
arises in a large variety of problems of 
mathematical physics\cite{2,3,4,5,6,7}  due to the conformal invariance 
of this
equation.
The general solution to (1.3) was obtained by Liouville 
\cite{8} .
It has the form
\begin{equation}
\varphi =\ln\frac{4\; A^{+'}(x^{+})A^{-'}(x^{-})}
{m^2 [A^{+}(x^{+})-\epsilon A^{-}(x^{-})]^2}.
\end{equation}
where  $x^{\pm}:= x^0 \pm x^1$ are
the `light-cone' coordinates, $A^{\pm}(x^{\pm})$ are 
any functions such that $A^{\pm '}>0$, 
$\epsilon$ is the sign of $R_0$ ($\epsilon := R_0/|R_0|$). 
Solution (1.4) is invariant under the
Bianchi\cite{9} transformations of $A^{\pm}$
\begin{equation}
A^+(x^+)\rightarrow \frac{aA^+(x^+)+b}{cA^+(x^+)+d},~~~~~
A^-(x^-)\rightarrow \frac{aA^-(x^-)+\epsilon b}{\epsilon cA^-(x^-)+d},
~~~~{\mbox {with}}~~~ad-bc=1.
\end{equation}

The aim of the present paper is to study the dynamics of a relativistic 
particle moving `freely' in the gravitational field (1.1), where 
$\varphi$ is a solution to Eq.(1.3).

Familiar action describing such a dynamical system is proportional to 
the lenght of a particle world-line, i.e., it is defined by
\begin{equation}
S= - m_0\int d\tau \sqrt{g_{\mu \nu}(X(\tau))\dot{x}^{\mu}(\tau)\dot{x}
^{\nu}(\tau}),
\end{equation}
where $\tau$ is an evolution parameter along trajectory $x^{\mu}(\tau),
 \;\dot{x}^{\mu}:= dx^{\mu}/d\tau$, and $m_0$ is a mass parameter.

The action is invariant under reparametrization $\tau 
\rightarrow f(\tau)$. This symmetry group leads to the constrained 
dynamics in the Hamiltonian formulation\cite{10,11,12}. 
The constraint is
\begin{equation}
\Phi := g^{\mu \nu}(X)p_{\mu}p_{ \nu}- m^2_0 = 0,
\end{equation}
 where $p_{\mu}:= \partial L/ \partial \dot{x}^{\mu}$ are 
 canonical momenta and $L$ is the Lagrangian corresponding to (1.6).

\noindent
The canonical Hamiltonian $H_c = (\partial L/\partial\dot{x}^{\nu})
\dot{x}^{\nu} - L$ vanishes. Making use of the Dirac procedure we 
obtain the Hamiltonian system 
\begin{equation}
\tilde{S} = \int p_{\mu}dx^{\mu} - \lambda (\tau)[g^{\mu \nu}(X)p_{\mu}
p_{\nu} - m^2_0]d{\tau},
\end{equation}
where $\lambda (\tau)$ is a Lagrange multiplier.

To investigate the gauge invariant system (1.8)
we apply the gauge-invariant quantization scheme \cite{13,14}.
We choose units such that $\hbar =1$.

The paper is organized as follows:

In Sec.II we present the method and technique for the investigation
of our system. We describe the general scheme of  
gauge-invariant 
Hamiltonian reduction
and give canonical quantization procedure 
which provides representation of the classical symmetry group. 
In cases this scheme fails, we consider another procedure based on 
the geometric quantization technique.

In Sec.III we study the classical dynamics of a particle for a simple 
singular Liouville field. The dynamics is essentially different for 
positive and negative curvatures of spacetime. In both cases we get:
(i) a complete set of dynamical integrals related to $SL(2,R)/Z_2$
symmetry, (ii) analytic description of particle trajectories, and (iii)
 Hamiltonian reduction in gauge-invariant variables. In the case of 
$R_0>0$ the reduced physical phase-space $\Gamma_{ph}$ is the 
coadjoint orbit of $SL(2,R)$ group with a positive `Casimir number' 
$C=2m_0/R_0$.
In the case of $R_0<0$ the physical phase space $\Gamma_{ph}$
is represented by a plane with canonical symplectic form and it
does not have the global
$SL(2,R)$ symmetry.

In Sec.IV we present quantization 
of our classical systems. For the positive curvature we 
arrive at the unitary irreducible representation of the universal 
covering group $\overline{SL(2,R)}$ 
which only for the discrete values of the 
Casimir number 
reproduces the classical $SL(2,R)/Z_2$ symmetry. Since 
the Casimir number is related to the parameter $m_0$, 
one gets 
the discretization of the particle mass $m_0$. 
In the case of the negative 
curvature the symmetry group generators are Hermitian operators 
which have non-unique self-adjoint extensions. Non-uniqueness is 
charecterized by the parameter $\alpha$ on the circle 
$z=\exp (i\alpha)$.
 Therefore, we have a continuous set of unitary non-equivalent 
 representations of $\overline{SL(2,R)}$. Only for 
 $z = 1$ it reproduces  
 $SL(2,R)/Z_2$ symmetry.

 In Sec.V we show that 
 the particle dynamics for an arbitrary Liouville field
 looks locally the same as for the examples considered
 in Sec.III. However, in general, $\Gamma_{ph}$ covers only a part
 of $SL(2,R)$ coadjoint orbit and sometimes the corresponding
 dynamical system is not well defined. This
 problem is related to the global symmetry properties
 of a spacetime manifold.
 For both signs of $R_0$ we give the examples of 
 manifolds free of this problem and describe the corresponding
 systems.

\section{A scheme of gauge invariant quantization}

We begin with a brief description of a Hamiltonian 
reduction scheme in gauge-invariant variables (see Refs. 13 and 14 
for details).

A gauge-invariant Lagrangian $L(Q,\dot Q)$ in Hamiltonian formalizm
leads to the extended phase space $\Gamma$ and the 
first class constraints.
Let $P:=(P_1, \ldots ,P_N)$ and $Q:=(Q^1,\dots ,Q^N)$ be coordinates on 
$\Gamma$, where there are $M$ first class constraints 
$\Phi _{\alpha}(P,Q)$ $(\alpha =1,...,M)$. The constraint surface
$\Gamma _c := \{(P,Q)\mid \Phi _{\alpha}(P,Q)=0;\;\;\alpha =1,\ldots ,
M\}$ has dimension $2N-M$. For such system there exist $2n\;\;( n=N-M)$
gauge-invariant functions $\xi ^a (P,Q)\;\; 
(a=1,\ldots , 2n)$, which are 
functionally independent on $\Gamma _c $. In general, there is no 
direct practical method for the construction of  ${\xi^a}'s$.
For some models one can guess them from the structure of gauge 
transformations. Another source is a global symmetry of $L(Q,\dot Q)$.
In particular, the dynamical integrals 
constructed by the Noether theorem are authomatically gauge invariant 
and if the symmetry group is large enough we can construct  all
$2n$  gauge invariant variables.

Suppose we know all $\xi ^a \;(a=1,...,2n)$. 
We complete them by $M$ variables $\eta^{\alpha}\;
(\alpha =1,\ldots ,M)$
to get a coordinate system $(\xi ^a,\eta ^{\alpha})$ on $
\Gamma _c$. The canonical Hamiltonian on $\Gamma _c$ is only 
a function of ${\xi ^a}'s$ since it is gauge invariant. 
The canonical 1-form $\theta :=P_{\mu}dQ^{\mu}$ on $\Gamma _c$ 
in $(\xi^a, \eta^\alpha )$ coordinates reads
\begin{equation}
{P_{\mu}dQ^{\mu}}_{\mid \Gamma _c} = \theta _a (\xi)d\xi ^a + 
dF(\xi ,\eta);\;\;\;\;a=1,\ldots ,2n;\;\;\;\;\mu =1,\ldots ,N,
\end{equation}
where $\xi :=(\xi ^1,...,\xi^{2n}),\;\;\eta :=(\eta ^1,\ldots ,
\eta ^M)$.
The first term in (2.1) $\;\theta _1 :=\theta _a(\xi)d\xi^a$ depends 
only on gauge invariant variables $\xi ^a$ and it defines a  1-form on 
the reduced physical phase space $\Gamma _{ph}$. The second term 
$\theta _2 := dF(\xi, \eta)$ can be neglected since it is an exact 
form (gives no contribution to the variation of an action integral).
The differential of the 1-form $\theta _1$ defines the symplectic form
on $\Gamma _{ph}$
\begin{equation}
\omega =\frac{1}{2}\omega _{ab}(\xi)d\xi ^a\wedge d\xi ^b \;\;\;\;
(a,b =1,\ldots ,2n),
\end{equation}
where $\omega _{ab}(\xi):= \partial _a\theta _b(\xi)-\partial _b
\theta _a(\xi) $. 

Calculation of the restricted 1-form ${P_{\mu}dQ^{\mu}}_
{\mid \Gamma _c}$
in $(\xi^a,\eta^\alpha)$ variables is an effective method for the
construction of a symplectic structure on $\Gamma _{ph}$. However,
in general, the described procedure has
only a local character and sometimes there are problems with the global
extension of the 1-form $\theta_1$.

Suppose ${\xi ^a}'s$ are dynamical integrals 
constructed by the 
Noether theorem using the symmetry group of the Lagrangian 
$L(Q,\dot{Q})$. 
Then the Poisson brackets of ${\xi ^a}'s$ (defined by the 
symplectic form $\omega$) coincide 
with the commutation relations of the corresponding symmetry group 
Lie algebra. Consistent quantization of such a system should give 
representation of these commutation relations on the appropriate 
Hilbert space.

To apply the 
canonical quantization method one should introduce 
new coordinates $p:=(p_1,\ldots ,p_n),~ q:=(q^1,\ldots ,q^n)$
on $\Gamma _{ph}$ such that (2.2) takes the  canonical form 
$\omega = dp_k\wedge dq^k\;\;(k=1,\dots ,n)$.

Standard canonical quantization implies the 
following prescription: $q^k\rightarrow \hat{q}^k:=q^k,\;\;p_k
\rightarrow \hat{p}_k := - i\partial_{q^k}$ and $\xi ^a(q,p)
\rightarrow \hat{\xi}^a:=\xi ^a(\hat{q},\hat{p}).$ In general, however,
there exists an operator ordering ambiguity; one does not know which 
ordering of operators $\hat{q},\hat{p}$ in $\hat{\xi}^a = \xi ^a(\hat
{q},\hat{p})$ reproduces the classical commutation 
relations of ${\xi ^a}'s$.
To investigate this problem we consider the following equations for 
$q^k=q^k(\xi)$:
\begin{equation}
\{\{\xi ^a,q^k\},q^l\}=0,\;\;\;\{q^k,q^l\}=0,\;\;\;(k,l=1,\ldots ,n),
\end{equation}
where ${\xi ^a}'s$ are gauge-invariant varibles constructed from 
the symmetry of the Lagrangian.

Suppose $q^k=q^k (\xi)\;\;(k=1,\ldots ,n)$ are the solutions of (2.3). 
We complete them by canonical conjugated `momenta' $p_k=p_k(\xi)\;\;
(k=1,\ldots ,n)$. In this case $\xi ^a\in 
\tilde{\cal O}(\Gamma_{ph}\mid q,p)$,
where $\tilde{\cal{O}}(\Gamma_{ph}\mid q,p)$ 
is the set of observables of 
the form
\begin{equation}
f(q,p)=p_kA^k(q)+U(q),
\end{equation}
with arbitrary functions ${A^k}$ and $U$ of only $q$ coordinates. 
It is 
clear that $\tilde{\cal{O}}(\Gamma _{ph}\mid q,p)$ 
is closed with respect to the Poisson 
bracket.

We define the quantum observables $\hat{f}$ corresponding to $f\in 
\tilde{\cal{O}}(\Gamma_{ph}\mid q,p)$ by 
\begin{equation}
\hat{f}:= \frac{1}{2}[A^k(q)\hat{p}_k+\hat{p}_kA^k(q)] + U(q) =
- iA^k(q)\partial_{q^k} - \frac{i}{2}\partial _{q^k}A^k(q)
+U(q).
\end{equation}
One can easily verify that $\hat{f}$ is Hermitian and for $f,g \in
\tilde{\cal O}(\Gamma _{ph}\mid q,p)$ we have
\begin{equation}
[\hat{f},\hat{g}] = - i\widehat{\{f,g\}}.
\end{equation}
Therefore, if we are able to solve (2.3) we get the quantum
realization of the
classical commutation relations. 

Described quantization scheme gives the prescription for the
construction of canonical coordinates convenient for the quantum
representation of a given classical symmetry. But it
may happen that this procedure does not lead to a global canonical
structure in $(p,q)$ coordinates. 
In that case the described method fails.
Sometimes one can find the global canonical coordinates $(p,q)$ but the
corresponding quantum observables (2.5) are only Hermitian and the
extension to self-adjoint operators is non-unique, or even does not
exist \cite{15} (see Sec. IV).

It can also happen that
(2.3) has a solution only for complex valued functions $q^k=q^k(\xi)$.
In that case
we propose the following scheme.
We introduce the Hilbert space $L_2(\Gamma _{ph})$ with the scalar
product
\begin{equation}
<\psi _2\mid\psi _1>:= \int \frac{d^{2n}\xi}{(2\pi )^n}\sqrt
{\omega (\xi)}\;\psi ^*_2(\xi)\psi _1(\xi),
\end{equation}
where $\omega (\xi)=det[\omega _{ab}(\xi)]\;\;$ (see (2.2)).
Obviously such a Hilbert space is too large for our system. 
We define a physical Hilbert space ${\cal H}_{ph}(q^1,\ldots ,q^n)$ 
as a subspace of $L_2(\Gamma _{ph})$ formed by the solutions to the 
equation
\begin{equation}
[iV_{q^k}+\theta (V_{q^k})] \psi (\xi)=0,\;\;\;(k=1,\ldots ,n),
\end{equation}
where 
\begin{equation}
V_{q^k}:= \omega ^{ab}(\xi)\partial _b q^k\partial_
{\xi ^a}:=V^a_{q^k}\partial_{\xi ^a},\;\;\;\;~~~
\theta (V_{q^k}):=\theta _a(\xi)V^a_{q^k}
(\xi ),
\end{equation}
and $\omega ^{ab}$ is the inverse to the symplectic matrix (2.2)
$\;(\omega _{ac}\omega ^{cb}=\delta ^b_a)$.
It is clear that if $q^k$ is a complex valued solution of (2.3),
then $q^{k*}$ is a solution as well. It turns out that when (2.8)
has solutions in $L_2(\Gamma_{ph})$ then the solutions of the
corresponding equation for $q^{k*}$ are not square integrable on
$\Gamma_{ph}$ and vice-versa.

Suppose that Eq.(2.8) has the solution in $L_2(\Gamma _{ph})$. We
introduce the prequantization operator ${\hat{\xi}}^a$ 
defined by \cite{16}
\begin{equation}
{\hat{\xi}}^a:=\xi ^a -\theta(V_{\xi ^a})- iV_{\xi ^a},
\end{equation}
where $V_{\xi ^a}$ is the Hamiltonian vector field corresponding 
to the observable ${\xi ^a}$ (see (2.9)).

The prequantization operators ${\hat{\xi}}^a$ are Hermitian on 
$L_2(\Gamma_{ph})$ and satisfy the relations \cite{16}
\begin{equation}
[{\hat{\xi}}^a,{\hat{\xi}}^b]= - i\widehat{\{\xi ^a, \xi ^b\}}.
\end{equation}
One can show \cite{17} that ${\cal H}_{ph}(q^1,\ldots ,q^n)$ is 
invariant under the action of ${\hat{\xi}}^a$ 
if (2.3) is fulfield. Therefore we arrive at the representation
of the classical commutation relations on 
${\cal H}_{ph}(q^1,\ldots ,q^n)$.

To illustrate the method, we consider a relativistic particle
in 2d Minkowski space with the Poincare group of symmetry. The
corresponding action is given by (1.6), where now $g_{\mu \nu}
 = diag (+1,-1)$.  
 On the constraint surface (1.7) we get 
\begin{equation}
 p_0 =-E({p_1}) := - \sqrt{(p_1)^2+m_0^2}
\end{equation}
(we assume that $\dot{x}^0(\tau)>0$). Restriction of the canonical
1-form $\theta\;$ on $\Gamma_c$ gives
\begin{equation}
\theta _{\mid \Phi =0} = p_1dx^1-E({p_1})dx^0,
\end{equation}

The infinitesimal Poincare transformations in 2-dimensions are
 \[   x^{\mu}\longrightarrow x^{\mu}+a^{\mu} +
b \varepsilon^{\mu \nu}
x_{\nu},  \]
where $ \varepsilon ^{\mu\nu}$ is an atisymmetric tensor with
$\varepsilon ^{01}=1,\;\;a^{\mu}$ and $b$ are group
parameters.
The corresponding three integrals
\begin{equation}
B:=p_{\mu}\varepsilon ^{\mu\nu}x_{\nu},\;\;\;\;P_{\mu}:=p_{\mu}\;\;\;
(\mu ,\nu =0,1)
 \end{equation}
 are gauge invariant since they have zero Poisson brackets with the
constraint $\phi =p_0^2-p_1^2-m_0^2$. On the  constraint surface
two integrals in (2.14) are independent. We choose them to be
$P:=P_1$ and $B$. By (2.12) and (2.14) we get
$x^1 =(B+P x^0)/{E({P})}$ (which describes the set of classical 
trajectories)
and (2.13) reads
\begin{equation}
\theta_{\mid \Phi =0} =
P d\Bigl(\frac{B}{E(P)}\Bigr)-
d\Bigl(\frac{m_0^2x^0}{E({P})}\Bigr).
\end{equation}
Therefore, the symplectic 2-form $\omega :=d\theta $
in $B,P$ variables reads
\begin{equation}
\omega =\frac{1}{{E}({P})}dP \wedge dB.
\end{equation}

According to (2.3) we consider the equtions for the coordinate $q$ 
\begin{equation}
\{\{E(P),q\},q\}=0,\;\;\;\{\{P,q\},q\}=0,\;\;\;
\{\{B,q\},q\}=0.
\end{equation}
We find the solution $q=P$, and the conjugated `momentum' is
$p= - B/{{E}({P})}.$

The Hilbert space ${\cal H}_{ph}$ is the space of square 
integrable functions $\psi (P)$ with the scalar product 
\[ <\psi _2\mid\psi _1> = \int dP~ \psi^*_2(P)\psi _1(P). \]
The quantum operators for the Poincare group generators are 
(see Eq.(2.5))
\[ {\hat{P}}_0={E}({P}),\;\;\;{\hat{P}}_1=P,\;\;\;
\hat{B}=i{E}({P})\frac{\partial}{\partial P}+\frac{i}
{2}\frac{P}{{E}({P})} . \]

It is easy to see that the quantum system we obtained is unitarily 
equivalent to the standard theory of a relativistic particle 
with the scalar product 
\[ <\phi _2\mid \phi _1> = \int \frac{dP}{{E}({P})}~ \phi ^*_2
(P)\phi _1(P), \]
where the unitary map is given by
$\phi (P) = \sqrt{{E}({P})}\;\psi (P)$. 

\section{Classical dynamics of a particle in the Liouville field}

The conformal transformation for the metric tensor (1.1) is defined by
\begin{eqnarray}
x^{\pm}\longrightarrow y^{\pm}(x^{\pm}),~~~~~~~~~~~~~~~~~~~~~~~~~~~~~~~
\nonumber \\
\varphi (x^{+},x^{-})\longrightarrow \tilde {\varphi} (x^{+},x^{-}):=
\varphi (y^{+}(x^{+}),y^{-}(x^{-}))+ 
\ln [y^{+'}(x^{+})y^{-'}(x^{-}) ].
\end{eqnarray} 
( We take both $y^{\pm'}$ to be positive to keep time direction fixed).

One can easily check that if $\varphi$ satisfies the Liouville 
equation (1.3) then $\tilde{\varphi}$ is the solution of (1.3) as well.
Therefore, making use of (3.1) one can generate a set of solutions 
of (1.3) starting from a known solution.  Simple solutions to (1.3) are 
\begin{eqnarray}
{\varphi}_+ = 
-2\ln (m|x|)=-2\ln (m|x^+-x^-|/2)\;\;\;&\mbox{for   $R_0>0$,}
\nonumber \\
{\varphi}_- =
-2\ln (m|t|)=-2\ln (m|x^++x^-|/2)\;\;\;&\mbox{for   $R_0<0$,}
\end{eqnarray}
where $m:=(\mid R_0\mid /2)^{1/2},\;\;t:=x^0,\;\;x:=x^1.$

Applying the conformal transformation (3.1) to the solution (3.2) 
we get the general solution (1.4) for $A^{\pm}(x^{\pm})=
y^{\pm}(x^{\pm})$. 
Therefore, the Liouville fields (with fixed $R_0$) 
are related to ${\varphi}_+$ (or ${\varphi}_-$) 
by the conformal transformations. 
In general, this is true only locally, since for an arbitrary functions
$A^{\pm}(x^\pm )$ the corresponding map (3.1) with 
$y^\pm =A^{\pm}(x^\pm )$ may be not well defined globally.
The action (1.5) is invariant with respect to any coordinate 
transformation. Thus, 
particle dynamics for an arbitrary 
Liouville field looks locally the same as for (3.2).

In this section we investigate the classical dynamics of a particle
in the Liouville fields given by (3.2). 
We will follow the scheme used in Sec.II for the particle
dynamics in the Minkowski space.
The coordinates $t$ and $x$
of the corresponding systems
are interpreted as the time and space coordinates, respectively,
and we consider only time-like trajectories 
with monotonically increasing time
$({\dot {x}}^{\pm}>0)$.
The time-independent field $\varphi_+$
leads to the conservation of particle energy, while homogeneity
of $\varphi_-$  in space provides the conservation of particle momentum.

\subsection{The case of positive curvature, $R_0>0$.}

The Lagrangian for the field $\varphi_+$ reads (see (1.6) and(3.2))
\begin{equation}
L=-2c \sqrt{\frac{\dot x^+\dot x^-}{(x^+-x^-)^2}}
~~~~~~(c:=\frac{m_0}{m}).
\end{equation}
Since (3.3) is singular at $x=1/2 (x^+-x^-)=0$
we `remove' this line and consider two systems for $x>0$ and
$x<0$ separately. The action (3.3) is invariant under the following
infinitesimal transformations
\begin{equation}
a)~ x^{\pm}\rightarrow x^{\pm}+\alpha_0,~~~~
b)~ x^{\pm}\rightarrow x^{\pm}+{\alpha_1}x^{\pm},~~~~
c)~ x^{\pm}\rightarrow{x^{\pm}}+\alpha_2(x^{\pm})^2.
\end{equation}
The infinitesimal transformations a) and b) define
the global transformations of the half-plane  $x>0$
(similarly for $x<0$), while the global transformation generated
by c) has singular point even for the whole plane $(t,x)$.
The global transformations generated by (3.4)
can be unified in the fractional-linear transformations
\begin{equation}
x^{\pm}\rightarrow\frac{ax^{\pm}+b}{cx^{\pm}+d},~~~ \mbox {with}
~~~~ab-bc=1.
\end{equation}
One can find that this `symmetry' of the Lagrangian (3.3) is
related to the Bianchi invariance of the Liouville fields 
(see (1.4) and (1.5)).

The dynamical integrals corresponding to the symmetry transformations
(3.4) are 
\begin{equation}
E=-(p_++p_-),~~~~K=p_+x^++p_-x^-,~~~~L=-(p_+x^{+2}+p_-x^{-2}).
\end{equation}
Since we consider time-like trajectories ($\dot x^\pm >0$) the
momenta $p_\pm$ are negative and our choice of the signs in
(3.6) leads to the positive energy $E$ (it corresponds to time
translation).

The Poisson brackets of the dynamical integrals (3.6) give $sl(2,R)$
algebra
\begin{equation}
\{ E, K\}_*=E,~~~~\{ K, L\}_*=L,~~~~\{ E, L\}_*=2K,
\end{equation}
where the Poisson brackets $\{ \cdot ,\cdot \}_*$ are calculated in
the canonical coordinates ($p_\pm , x^\pm $) of the extended 
phase space
$\Gamma$.

The mass-shell condition (1.7) takes the form
\begin{equation}
\Phi :=m^2(x^+-x^-)^2p_+p_--m_0^2=0.
\end{equation}
The dynamical integrals (3.6) are gauge-invariant since they have
zero Poisson brackets with $\Phi$. On the constraint surface (3.8) 
we find the relation
\begin{equation}
EL-K^2=c^2.
\end{equation}
It defines the upper hyperboloid in $E, K, L$ space ($E>0,
L>0$), which is the coadjoint orbit of the group $SL(2,R)$.
We can choose $E$ and $K$ as the global coordinates on this
hyperboloid. Thus, the physical phase space $\Gamma_{ph}$
is given on the half-plane ($E>0,K$). But one should check whether
$\Gamma_{ph}$ covers the entire half-plane.

For $t=0$ Eq.(3.6) reads
\begin{equation}
E=\sqrt {P^2+c^2/x^2}, ~~~~K=Px,
\end{equation}
where $P:=p_+-p_-$. Eq.(3.10) gives the diffeomorphism of 
the half-planes ($x>0,P$) and 
($E>0,K$).
Therefore, all points of the half-plane ($E>0,K$) are available
for the system with $x>0$ (similarly for $x<0$).

Eqs.(3.6) and (3.9) define the trajectories
\begin{equation}
\left (x^++\frac{K}{E}\right )\left (x^-+\frac{K}{E}\right )
=-\frac{c^2}{E^2},
\end{equation}
which are hyperbolas with light-cone asymptotics. The particle
trajectory never crosses the singularity at $x=0$. 
Thus our two systems for 
$x>0$ and $x<0$ are independent.

We can accomplish the Hamiltonian reduction (2.1) using the 
gauge-invariant variables $E$ and $K$ as $\xi^a$ variables, and
$t$ as $\eta$ one. The result is
\begin{equation}
\theta_{\mid \Gamma_{c}}=p_+dx^++p_-dx^-_{\mid_{\Phi=0}}=
Ed\left (\frac{K}{E}\right )+\left (\frac{y}{y^2+c^2}-1\right )dy,
\end{equation}
where $y:=Et-K$. Respectively, the symplectic form on $\Gamma_{ph}$
is given by
\begin{equation}
\omega=\frac{dE\wedge dK}{E}.
\end{equation}
The symplectic form (3.13) reproduces the $sl(2,R)$ algebra
(3.7) for $E,K$ and $L=(K^2+c^2)/E$ observables. 

Since the canonical Hamiltonian of our system is equal to zero,
there is no dynamics on $\Gamma_{ph}$. Each point of $\Gamma_{ph}$
defines the trajectory given by (3.11). Thus $\Gamma_{ph}$ is
associated with the space of trajectories (space of motions \cite{16}).
This space has the global $SL(2,R)/Z_2$ symmetry, whereas
the corresponding `symmetry' transformations (3.5) for the 
Lagrangian (3.3) are not defined globally. 

\subsection{The case of negative curvature, $R_0<0$}

The Lagrangian for this case is (see $\varphi_-$ in (3.2)) 
\begin{equation}
L=-2c\sqrt{\frac{\dot x^+\dot x^-}{(x^++x^-)^2}},~~~~~~~
(c:=\frac{m_0}{m})
\end{equation}
There is a singularity at $t=1/2 (x^++x^-)=0$ and we consider
the dynamics of a particle for $t>0$ and $t<0$ separately.

The infinitesimal symmetry transformations are now 
\begin{equation}
x^{\pm}\rightarrow x^{\pm}\pm\alpha_0,~~~~~
x^{\pm}\rightarrow x^{\pm}\pm\alpha_1x^{\pm},~~~~~
x^{\pm}\rightarrow x^{\pm}\pm\alpha_2(x^{\pm})^2,
\end{equation}
which lead to the corresponding dynamical integrals
\begin{equation}
P=p_+-p_-,~~~~K=p_+x^++p_-x^-,~~~~M=p_+(x^+)^2-p_-(x^-)^2,
\end{equation}
with the commutation relations of $sl(2,R)$ algebra on the extended
phase space $\Gamma$
\begin{equation}
\{ P , K\}_*=P,~~~~\{ K , M\}_*=M,~~~~\{ P , M\}_*=2K.
\end{equation}
Taking into account the mass-shell conditions (1.7), now we get the
one-sheet hyperboloid
\begin{equation}
K^2-PM=c^2
\end{equation}
and the relations (3.16) read
\begin{equation}
K=Px-t\sqrt{P^2+(c/t)^2},~~~~M=P(t^2+x^2)-2tx\sqrt{P^2+(c/t)^2}.
\end{equation}

First, let us consider the system with $t>0$. One can see that 
$\Gamma_{ph}$ is a hyperboloid without the line defined by $P=0,
K=c$. Therefore, it is diffeomorphic to a plane $\bf {R}^2$. 
For $P=0$
Eq.(3.19) reads
\begin{equation}
K=-c,~~~~~~~x=-\frac{M}{2c},
\end{equation}
which describes a rest particle. For $P\neq0$ the trajectory is a
hyperbola with a light-cone asymptotics given by
\begin{equation}
x=\frac{K+\sqrt{P^2t^2+c^2}}{P}.
\end{equation}
The reduction procedure (2.1) in this case reads
\begin{equation}
\theta_{\mid \Gamma_{c}}=Pd\left (\frac{K+c}{P}\right )+
yd\left (\frac{y}{\sqrt{y^2+c^2}+c}\right )-c\frac{dt}{t},~~~~
\mbox {with}~~ y:=Pt.
 \end{equation}

Introducing the new variable $Q:={(K+c)}/{P}$ we get the canonical
symplectic form $\omega =dP\wedge dQ$. 
In the canonical coordinates ($P,Q$) (3.19) reads
\begin{equation}
K=PQ-c,~~~~~~~~M=PQ^2-2Qc.
\end{equation}
It is clear that ($P,Q$) plane represents $\Gamma_{ph}$.

For further consideration it is convenient to introduce the new set
of dynamical integrals $J_\alpha ~(\alpha =0,1,2)$
\begin{equation}
J_0:=\frac{1}{2}(P+M),~~~~~J_1:=\frac{1}{2}(P-M),~~~~~J_2:=K,
\end{equation}
with the commutation relations
\begin{equation}
\{ J_i, J_j\}=\varepsilon_{ijk}J^k,
\end{equation}
where $\varepsilon_{ijk}$ is a totally antisymmetric
tensor defined by $\varepsilon_{012}=1$, 
$J^k =\eta^{kl}
J_l$ (with $\eta^{kl}=diag(+,-,-))$. 

Similarly to the $R_0>0$ case, $\Gamma_{ph}$ is the space of
trajectories. But now, $SL(2,R)/Z_2$ is not a global symmetry on 
$\Gamma_{ph}$. The global symmetry transformations are 
generated only by $P$ and
$K$ obsevables. The same is true for the system with $t<0$.

In this case $\Gamma_{ph}$ is a hyperboloid without a line 
$P=0,~ K=-c$. Thus, $\Gamma_{ph}$ is again diffeomorphic to a plane.
The trajectories are now defined by
\begin{equation}
x=\frac{K-\sqrt{P^2t^2+c^2}}{P}~~~~\mbox{for}~~~P\neq 0,~~~~~~ 
x=\frac{M}{2c}~~~~\mbox {for}~~~ P=0.
\end{equation}
The reduced 1-form reads
\begin{equation}
\theta_{\mid \Gamma_{c}}=Pd\left (\frac{K-c}{P}\right )-
yd\left (\frac{y}{\sqrt{y^2+c^2}+y}\right )+c\frac{dt}{t},~~~
\mbox {with}~~~y=Pt.
\end{equation}
The coordinate canonically conjugated to $P$ is $Q:=(K-c)/P$.
Thus, $\omega =dP\wedge dQ$ and
\begin{equation}
K=PQ+c,~~~~~M=PQ^2+2cQ.
\end{equation}

One can consider both systems for $t>0$ and $t<0$
as two parts of a one system with $t$ (time) going continously from
negative to positive direction. In such  case one should find
the ways of glueing trajectories at $t=0$. This corresponds to the
identification of points of two physical phase spaces $\Gamma_{ph}$ for
$t<0$ and $\Gamma_{ph}$ for $t>0$. If we insist on the continuity
of trajectories at $t=0$, the dynamical integrals $P, K$ and $M$
change signs (compare (3.20)-(3.21) with (3.26)). 
There is another possibility
of identification of these two phase spaces by requiring the
conservation of the dynamical integrals $P,K$ and $M$. 
But this leads to the
discontinuity of trajectories (see again (3.20)-(3.21) and (3.26)).
In Sec.IV we show that the corresponding abmiguity in quantum case
is expressed by the unitary non-equivalent
quantizations.

Consideration of $(t,x)$ plane with singular Liouville 
field $\varphi$ leads to 
singular spacetime structure which is a subtle aspect of general 
relativity and we are going to discuss the related aspects of the
model elsewhere.

\section{Quantization}

For quantization of the obtained reduced  Hamiltonian systems
we apply the method described in Sec.II.

\subsection{The case $R_0>0$}

One can see that if $q$ is a solution of Eq.(2.3), then $f(q)$ is a
solution as well (where $f$  is a function of $q$). However, both
constructions (2.4) and (2.8) are invariant under this choice.
Using this freedom we look for the solution of (2.3) in the form
$q=A(E)B(K)$ (where $A$, $B$ are some functions). Then the solution 
of (2.3) for $\xi^1=E$, $\xi^2=K$ and $\xi^3=L$ fixes the functions 
$A$ and $B$, and we end up with complex valued functions
\begin{equation}
q=\frac{K+ic}{E},~~~~~~q^*=\frac{K-ic}{E}.
\end{equation}
The prequantization operators in ($q, q^*$) variables are (see (2.10))
\begin{equation}
\hat E=-i(\partial_q+\partial_{q^*}),~~~
\hat K=-i(q\partial_q+q^*\partial_{q^*}),~~~
\hat L=ic(q^*-q)-i(q^2\partial_q+q^{*2}\partial_{q^*}).
\end{equation}
For our choice of variables (see (4.1))
the physical Hilbert space ${\cal {H}}_{ph}(q)$ is defined by the
solutions to the equation (see(2.8))
\begin{equation}
\left (\partial_{q^*}+\frac{c}{q-q^*}\right )\Psi (q,q^*)=0.
\end{equation}
The general solutions of (4.3) is 
\begin{equation}
\Psi (q,q^*)=\left (\frac{q-q^*}{2i}\right )^c\psi(q),
\end{equation}
where $\psi (q)$ is a holomorphic function on the upper half-plane,
Im$q>0$ (see (4.1)).

The scalar product (2.7) takes the form
\[ <\psi _2\mid\psi _1> = \int\;\frac{dqdq^*}{2\pi}\left(\frac{q-q^*}
{2i}\right)^{2(c-1)}{\psi}_2^*(q)\psi_1(q), \]
where we should impose $c>1/2$ to avoid the non-integrable 
singularity at $q-q^*=0$.
The class of square-integrable functions is given by
\[ \psi(q)=(q+i)^{-2c}\;\sum_{n\geq 0}c_n\left(\frac{q-i}{q+i}
\right)^n,~~~ \mbox {with}~~~\sum_{n\geq 0}\mid c_n\mid ^2 < \infty .\]          
One can easily show that (4.2) and (4.4) define the following 
representation of $sl(2,R)$ algebra on the holomorphic functions $\psi
(q)$
\begin{equation}
\hat{E}\rightarrow -i\partial_q,~~~~\hat{K}\rightarrow -iq\partial_q
-ic,~~~~\hat{L}\rightarrow -iq^2\partial_q-2icq.
\end{equation}
The Casimir operator $\hat{C}:=(\hat{E}\hat{L}+\hat{L}\hat{E})/2-
\hat{K}^2$ for the representation (4.5) reads $\hat{C} = c(c-1)\hat
{\bf{1}}.$ We take $c>1$ to have positive Casimir number 
in quantum case  
as well (see (3.9)). Therefore, we arrive at the unitary irreducible 
representation of $sl(2,R)$ algebra \cite{18}.

Due to the simple form of the operators (4.5) the eigenvalue problem 
can be solved easily. The operators $\hat{E}$ and $\hat{L}$ have 
positive continuous spectra but the spectrum of the operator 
\begin{equation}
\hat{N}
:=(\hat{E}+\hat{L})/2
\end{equation}
is discrete with the eigenvalues $N_n=c+n\;( n=0,1,
\ldots)$ and the eigenfunctions 
$$\psi_n \sim (q+i)^{-2c}\left( \frac{q-i}{q+i}\right)^n.
$$
Irreducible representation of $sl(2,R)$ algebra by the operators (4.5) 
leads to the corresponding representation of the universal covering 
group $\overline{SL(2,R)}.$ For $c=k/2\;(k=3,4,\ldots)$ one gets 
the representation of $SL(2,R)$ group \cite{18}.
Similarly, for $c=l\;(l=2,3,
\ldots)$ we arrive at the representation of $SL(2,R)/Z_2$ group. 
Therefore, the requirement that the quantum system reproduces 
the classical $SL(2,R)/Z_2$ symmetry yields that $c$ is discrete, 
which means that the mass of a particle $m_0$ is discrete ($c:=m_0/m,
\;m$ is a parameter of the Liouville equation).

\subsection{The case $R_0<0$}

Since now our reduced phase space coordinates $(P,Q)$ are canonical 
and the symmetry group generators (3.23) and (3.24) 
belong to $\tilde {\cal{O}}(\Gamma_{ph}\mid P,Q)$ we can apply 
the procedure (2.5). We get the quantum observables
\begin{eqnarray}
\hat{J}_0=-\frac{i}{2}(1+Q^2)\partial_Q-(c+\frac{i}{2})Q,~~~~
\hat{J}_1=-\frac{i}{2}(1-Q^2)\partial_Q+(c+\frac{i}{2})Q,\nonumber \\
\hat{J}_2=-iQ\partial_Q-(c+\frac{i}{2}).~~~~~~~~~~~~~~~~~~~~~~~~~~~~
\end{eqnarray}
These generators are Hermitian on $L_2(\bf {R})$ 
and give the representation 
of $sl(2,R)$ algebra (3.17) \cite{18}. However, according to the 
quantization principles quantum observables should be represented by 
self-adjoint operators\cite{}. Therefore, one should find an extension 
of operators $\hat{P},\hat{K}$ and $\hat{M}$ to self-adjoint ones. 
As was mensioned in Sec.II sometimes such an extension is not unique. 
For example\cite{15}, the operator
\begin{equation}
\hat{T}:=-i\frac{d}{dx}
\end{equation}
is an unbounded Hermitian operator on the Hilbert space $L_2([0,2\pi])$
 with the domain
\[ D_{\hat{T}}:=\{\psi (x)\mid\psi(0)=0=\psi(2\pi)\}. \]
The operator $\hat{T}$ has a set of unitary non-equivalent 
self-adjoint  extensions $\hat{T}_\alpha$ paremetrized by $\alpha \in 
S^1$. The domain of $\hat{T}_\alpha$ is
\[ D_{\hat{T}_{\alpha}}:=\{\psi (x)\mid\psi(2\pi )=
\exp(i\alpha)\psi(0)\}. \]
The eigenvalue problem for $\hat T_\alpha$ reads
\begin{equation}
\hat{T}_\alpha\psi_{\lambda_n}(x)=
\lambda_n(\alpha)\psi_{\lambda_n}(x),
\end{equation}
where
\begin{equation}
\psi_{\lambda_n}(x):=(2\pi)^{-1/2}\exp\{i\lambda_n(\alpha)x\};~~~~
\lambda_n(\alpha):=\frac{\alpha}{2\pi}+n;~~~~n=0,\pm1,\ldots
\end{equation}

It turns out that the self-adjoint extension of the operators (4.7) 
is non-unique as well. To show this we consider the unitary 
transformation $U:L_2({\bf{R}})\rightarrow L_2([0,2\pi])$ defined by 
\[ U\Psi (Q)=\psi (x):=\frac{\exp{(-2ic\ln \sin{(x/2)})}}
{2\sin^2{(x/2)}}\;
\Psi
(-\cot{(x/2)}),\]
where $Q\in {\bf {R}},~~x\in ]0,2\pi[;~~\Psi(Q)\in L_2({\bf{R}}),~~
\psi (x)\in L_2([0,2\pi])$.

Transformed operators (4.7) read
\begin{equation}
\hat{J}_0=-i\partial_x,~~~~\hat{J}_1=i\cos x\partial_x -(c+i/2)\sin x,
~~~~\hat{J}_2=i\sin x\partial_x+(c+i/2)\cos x.
\end{equation}
$\hat{J}_0$ is the operator (4.8). Thus, it has a set of 
self-adjoint extensions $\hat{J}_{0\alpha}=\hat{T}_\alpha$. Since 
$\sin x$ and $\cos x$ in (4.11) are bounded operators and $D_{\hat{T}
_\alpha}$ is invariant under their action, the self-adjoint extensions 
of $\hat{J}_1$ and $\hat{J}_2$ read
\begin{equation}
\hat{J}_{1\alpha}=-\cos x~ \hat{T}_\alpha 
-(c+i/2)\sin x,~~~~
\hat{J}_{2\alpha}=-\sin x~ \hat{T}_\alpha 
+(c+i/2)\cos x.
\end{equation}
This way we arrive at the unitary non-equivalent representations
(parametrized by $\alpha\in S^1$)
of the universal covering group $\overline{SL(2,R)}$ \cite{}.
The Casimir number for all these representations is $C=-(c^2+1/4)$.

\section{General Liouville field and the global $SL(2,R)/Z_2$ symmetry}

For the general solution (1.4) the Lagrangian of (1.6) has the form
\begin{equation}
L=-2c\sqrt{\frac{\dot A^+(x^+(\tau ))\dot A^-(x^-(\tau ))}
{[A^+(x^+(\tau ))-\epsilon A^-(x^-(\tau ))]^2}},~~~~~~~
(c:=\frac{m_0}{m}).
\end{equation}
Choosing the coordinates $y^{\pm}=A^{\pm}(x^{\pm})$ we end up with
(3.3) or (3.14) depanding on the signature of $\epsilon$.
The Bianchi invariance (1.5) leads to the symmetry of the
Lagrangian (5.1). In $y^{\pm}$ coordinates symmetry transformations
read
\begin{equation}
 y^+\rightarrow \frac{ay^++b}{cy^++d},~~~~~
 y^-\rightarrow \frac{ay^-+\epsilon b}{\epsilon cy^-+d},
 ~~~~{\mbox {with}}~~~ad-bc=1.
\end{equation}
Therefore, one can repeat the procedure of Sec.III for the general
case and obtain the reduced phase space $\Gamma_{ph}$ with 
the symplectic form on the corresponding coadjoint orbit of
$SL(2,R)$ group.
However, one should take into account 
that $\Gamma_{ph}$, in general, covers only a part of the 
hyperboloid.

If the Liouville field (1.4) is not singular \cite{20,21} in
${\bf{R}}^2$ we have $A^+(x^+)-\epsilon A^-(x^-)>0$. Since
$A^+(\cdot )$ and $A^-(\cdot )$ are the functions of different
arguments they must be bounded. For the 
bounded coordinates $y^{\pm}=A^{\pm}(x^{\pm})$ the 
symmetry transformations (5.2) are not well defined.  
In the case of singular Liouville field  singularities 
are given by 
$ A^+(x^+)-\epsilon A^-(x^-)=0,$
which define \cite{22} non-intersecting time-like 
lines for $\epsilon >0$
and non-intersecting space-like lines for $\epsilon <0$.
The Lagrangian is well defined on some domain bounded
by two lines of singularities. In general, the symmetry
transformations (5.2) for such a domain again are 
not defined globally.
If the Lagrangian (5.1) does not have the global $SL(2,R)/Z_2$ symmetry,
it can happen that this symmetry is absent
from $\Gamma_{ph}$ as well. In that case 
$\Gamma_{ph}$ is only a part of the coadjoint orbit.
It is clear that the dynamical system given on a part of the 
hyperboloid 
may be not well defined. 

The domain of  ${\bf{R}}^2$ with non-singular Liouville 
field can be considered as a patch of a new spacetime manifold
with a constant curvature $R_0$. If the new manifold has global
$SL(2,R)/Z_2$ symmetry, the system is well defined. 

An example of such a system is a particle 
in the Liouville field
\begin{equation}
 \varphi =-\ln{\sin^2\rho}
 \end{equation}
given on 
a cylinder ${\bf {C}}:=\{(\sigma ,\rho )|
~\sigma\in S^1,~\rho\in ]0,\pi [~\}$, 
where we use `dimensionless' coordinates $\rho :=mt, \sigma :=mx$.
The corresponding pseudo-Riemannian manifold has
the negative curvature ($R_0=-2m^2$). 
The Lagrangian of the system is
\begin{equation}
L=-c\sqrt{\frac{\dot \rho^2-\dot\sigma^2}{\sin^2\rho}}~~~~~~(c:=
\frac{m_o}{m}) 
\end{equation}
We again assume that the trajectories are time-like $(|\dot \rho |>
|\dot\sigma |)$ and $\dot \rho >0$.

The infinitesimal symmetry transformations are
\begin{eqnarray}
a)~ \rho\rightarrow \rho,~~ \sigma\rightarrow \sigma +\alpha_0;~~~~~~
b)~ \rho\rightarrow \rho +\alpha_1 \sin \rho\cos\sigma,~~
\sigma\rightarrow \sigma +\alpha_1\cos\rho\sin\sigma ;\nonumber \\
c)~\rho \rightarrow \rho +\alpha_2 \sin \rho\sin\sigma,~~ 
\sigma\rightarrow \sigma -\alpha_2\cos\rho\cos\sigma ;~~~~~~~~~~~~~~~~~ 
\end{eqnarray}
which give well-defined global action of the group $SL(2,R)/Z_2$
on the cylinder $\bf C$.

The dynamical integrals corresponding to (5.5) read 
\begin{equation}
J_0=p_\sigma ,~~~J_1=p_\rho\sin \rho\cos\sigma +
p_\sigma\cos\rho\sin\sigma ,~~~ J_2=p_\rho\sin \rho\sin\sigma
-p_\sigma\cos\rho\cos\sigma,
\end{equation}
where $p_\sigma :=\partial L/\partial \dot\sigma$ and
$p_\rho:=\partial L/\partial \dot\rho$ are canonical momenta.
They satisfy the commutation relations (3.25)
on the extended phase space $\Gamma$.
The mass-shell condition (1.7)
leads to the one-sheet hyperboloid 
\begin{equation}
J_kJ^k=-c^2.
\end{equation}
Eq.(5.6) and (5.7) define particle  
trajectories on the cylinder.
For $\rho =\pi /2$ (5.6) and (5.7) give
$$
J_0=p_\sigma,~~~~~~~~J_1=-\sqrt{p_\sigma^2+c^2}\cos\sigma,~~~~~~
~~ J_2=-\sqrt{p_\sigma^2+c^2}\sin\sigma .
$$
Therefore, the physical phase space covers the entire hyperboloid
(5.7), which has the global $SL(2,R)/Z_2$ symmetry.

The Hamiltonian reduction procedure (2.1) gives the symplectic
form $\omega =dJ \wedge d\phi$ and the generators (5.6) 
read
\begin{equation}
J_0=J,~~~~~~~~J_1=-\sqrt{J^2+c^2}\cos\phi,~~~~~~
~~ J_2=-\sqrt{J^2+c^2}\sin\phi .
\end{equation}

The reduced physical phase space $\Gamma_{ph}$ has a global canonical
structure, since $\Gamma_{ph}$ is isomorphic to $T^*S^1$.
Thus, we can use the canonical quantization rule 
$\hat J=-i\partial_\phi$. However, there is an operator ordering 
problem for $\hat J_1$ and $\hat J_2$. Therefore, we apply the method
of Sec.II. The solution of (2.3) for $\xi^1=J,~\xi^2=\sqrt{J^2+c^2}
\cos\phi$ and $\xi^3=\sqrt{J^2+c^2}\sin\phi$
gives the cyclic coordinate $q\in S^1$, where
\begin{equation}
q=\phi +\phi_0 ~~~~~~{\mbox {with}}~~~~~~\cot\phi_0=J/c~~~~~
~~~~(\phi_0\in ]0,\pi [).
\end{equation}
$J$ is the canonically conjugated momentum to $q$. 
In the new coordinates 
$(J,q)$ the dynamical integrals (5.6) are linear in $J$
$$
J_0=J,~~~~~~~J_1=-J\cos q-c\sin q,~~~~~~J_2=-J\sin q+c\cos q.
$$
Using (2.5) we arrive at the operators
\begin{equation}
\hat J_0=-i\partial_q,~~~~~~~\hat J_1=i\cos q\partial_q-(c+i/2)
\sin q,~~~~~~\hat J_2=i\sin q\partial_q+(c+i/2)\cos q.
\end{equation}
The operators (5.10) are self-adjoint on the Hilbert space $L_2(S^1)$
with the scalar product
$$
<\psi_2~|~\psi_1>:=\int_0^{2\pi} dq~\psi^*_2(q)\psi_1(q).
$$
They define an irreducible unitary representation of $SL(2,R)/Z_2$
group \cite{18} with the Casimir number $C=-(c^2+1/4)$.

Comparing (5.10) and (4.12), we see that for $\alpha =0$
these representations are unitarily equivalent.

\vspace{5mm}

In the case of positive curvature ($R_0=2m^2$) one can consider 
the Liouville field
\begin{equation}
\varphi =-\ln \sin^2\sigma ,
\end{equation}
where we again use the dimensionless coordinates $\rho =mt,\sigma =mx$.
Taking `unbounded' time coordinate $ \rho\in {\bf{R}}$ we get the
Liouville field 
on the stripe ${\bf{S}}:=\{(\sigma ,\rho )|~\sigma\in ]0,\pi [,~
\rho\in {\bf{R}}\}$. The corresponding Lagrangian has
the global $\overline{SL(2,R)}$ symmetry. 
The time independent field (5.11) leads to the conservation of energy.
This system describes the periodic motion of a particle between
two singularities at $\sigma =0$ and $\sigma =\pi$.
The reduced phase space is
the coadjoint orbit of $SL(2,R)$.
The quantum operator for the energy coincides 
with the generator (4.6) 
and has the same discrete spectrum.

\acknowledgments

One of us (G.J) thanks M. Maziashvili and A. Pogrebkov for helpful
discusions.
This work was supported by the grants from:
INTAS, RFBR (96-01-00344),
the Georgian Academy of Sciences, the
Polish Academy of Sciences, and the 
So{\l}tan Institute for Nuclear Studies.

\nopagebreak

\end{document}